\documentstyle[12pt]{article}
\begin{document}
\begin{titlepage}
\begin{center}
{\Large \bf  Finiteness of Soft Terms in \protect\\[0.5cm]
Finite N=1  SUSY Gauge Theories}

\vglue 10mm
{\bf  D.I.~Kazakov }

\vglue 5mm
{\it Bogoliubov Laboratory of Theoretical Physics, \\
Joint Institute for Nuclear Research, \\
141 980 Dubna, Moscow Region, RUSSIA\\
e-mail: kazakovd@thsun1.jinr.dubna.su}
\end{center}

\vglue 20mm
\begin{abstract}
Recently proposed relations between the renormalization group functions
of the soft supersymmetry breaking terms and those of a rigid theory
allow one to consider a possibility of constructing a totally
all loop finite N=1 SUSY gauge theory, including the soft SUSY breaking
terms.  The requirement of finiteness, which can be satisfied in
previously constructed finite SUSY GUTs, imposes some constraints on
the SUSY breaking parameters which, in the leading order, coincide with
those originating from the supergravity and superstring-inspired
models. Explicit relations between the soft terms, which lead to a
completely finite theory in any loop order, are given.
\end{abstract}

\vglue 10mm
\noindent PACS numbers:
11.10Gh, 11.10Hi, 11.30Pb

\end{titlepage}

\section{Introduction}

In the recent papers~\cite{NewJ, AKK}, the relations between the soft
term renormalizations and those of an unbroken SUSY gauge theory have been
derived. It has been shown that the soft term renormalizations are not
independent, but can be calculated from the known renormalizations of a rigid
theory with the help of the differential operators. The explicit form
of these operators has been found in a general case and in some
particular models like SUSY GUTs or the MSSM.

As an application of the above mentioned relations we consider a
possibility of constructing totally finite supersymmetric theories
including the soft breaking terms. This problem has already been
discussed several times~\cite{Mezinchescu, Jones}. In particular, it
has been found that imposing some condition on the soft terms (see
below) one can reach complete one and two-loop finiteness. This is
similar to the finiteness of the rigid SUSY theories themselves where
imposing conditions on the particle content and the Yukawa couplings
one can reach finiteness in one and two loops~\cite{Finite}.  To go
further, one has to fine-tune the Yukawa couplings order by order in
perturbation theory~\cite{EKT, Jones1, Sibold}.  We  will show below
that the same procedure works for the softly broken theory as well. By
choosing the soft terms in a proper way one can reach complete all loop
finiteness.  Moreover, there is no new fine-tuning.  The soft terms are
fine-tuned in exactly the same way as the corresponding Yukawa
couplings.

\section{Soft SUSY Breaking  and Renormalization}

Consider an arbitrary $N=1$ SUSY gauge theory with unbroken SUSY. The
Lagrangian of a rigid theory is given by
\begin{eqnarray}
{\cal L}_{rigid} &=& \int d^2\theta~\frac{1}{4g^2}{\rm Tr}W^{\alpha}W_{\alpha}
+ \int d^2\bar{\theta}~\frac{1}{4g^2}{\rm Tr}
\bar{W}^{\alpha}\bar{W}_{\alpha}.  \label{rigidlag} \\
&+& \int d^2\theta  d^2\bar{\theta} ~~\bar{\Phi}^i (e^{V})^j_i\Phi_j + \int
 d^2\theta ~~{\cal W} + \int d^2\bar{\theta} ~~\bar{\cal W},  \nonumber
\end{eqnarray}
where $W^{\alpha}$ is the field strength chiral superfield
defined by
\begin{eqnarray}
W^{\alpha} = \overline{D}^2 \left( e^{-V} D^{\alpha}e^V \right), \nonumber
\end{eqnarray}
and the superpotential ${\cal W}$  has the form
\begin{equation}
{\cal  W}=\frac{1}{6}\lambda^{ijk}\Phi_i\Phi_j\Phi_k +\frac{1}{2}
M^{ij}\Phi_i\Phi_j.\label{rigid}
\end{equation}

To  perform the SUSY breaking, which satisfies the requirement of "softness",
one can introduce a gaugino mass term as well as cubic and quadratic
interactions of the scalar superpartners of the matter fields.
They should not break a gauge invariance.

The soft SUSY breaking term satisfying these requirements can be written
as~\cite{spurion}
\begin{eqnarray}
-{\cal L}_{soft-breaking} &=& \frac{m_A}{2}\lambda\lambda +
\frac{m_A}{2}\bar \lambda \bar \lambda, \nonumber  \\
&+&\left[\frac 16 A^{ijk}
\phi_i\phi_j\phi_k+ \frac 12 B^{ij}\phi_i\phi_j +h.c.\right]
+(m^2)^i_j\phi^{*}_i\phi^j,\label{sofl}
\end{eqnarray}
where $\lambda$ is  the gaugino field and $\phi_i$ is the lower
component of the chiral matter superfield.

The key point in establishing the relation between the soft term
renormalizations and those of a rigid theory is the
possibility to rewrite the soft terms in terms of
superfields~\cite{spurion}.  To do this, let us  introduce the
external spurion superfields $\eta=\theta^2$ and $\bar
\eta=\bar{\theta}^2$, where $\theta$ and $\bar \theta$ are the
grassmannian parameters~\cite{supergraph}.  The softly broken
Lagrangian can then be written as~\cite{Yamada} \begin{eqnarray} {\cal
L}_{soft} &=& \int d^2\theta~\frac{1}{4g^2}(1-2\mu\theta^2) {\rm
Tr}W^{\alpha}W_{\alpha} + \int
 d^2\bar{\theta}~\frac{1}{4g^2}(1-2\bar{\mu}\bar{\theta}^2) {\rm
Tr}\bar{W}^{\alpha}\bar{W}_{\alpha}.  \nonumber \\
&+&\int d^2\theta
d^2\bar{\theta} ~~\bar{\Phi}^i(\delta^k_i -(m^2)^k_i\eta
\bar{\eta})(e^V)^j_k\Phi_j   \label{ssofl2} \\
&+& \int  d^2\theta
\left[\frac 16 (\lambda^{ijk}-A^{ijk} \eta)\Phi_i\Phi_j\Phi_k+ \frac 12
(M^{ij}-B^{ij}\eta ) \Phi_i\Phi_j \right] +h.c.  \nonumber
\end{eqnarray}
The external spurion superfield can be considered as a vacuum
expectation value of a dilaton superfield emerging from supergravity,
 however, we will not explore this fact in further discussion.

As has been shown in Ref.~\cite{AKK} the following statement is valid:

\newtheorem{statement}{Statement}
\begin{statement}
{\it Let a rigid theory
(\ref{rigidlag}, \ref{rigid}) be renormalized via introduction of the
renormalization constants $Z_i$, defined within some minimal subtraction
massless scheme. Then, a softly broken theory (\ref{ssofl2})
is renormalized via introduction of the renormalization superfields
$\tilde{Z}_i$ which are related to $Z_i$ by the coupling constants
redefinition
\begin{equation}
\tilde{Z}_i(g,\lambda ,\bar \lambda)
 =Z_i(\tilde{g}^2,\tilde{\lambda},\tilde{\bar \lambda}),
\label{Z}
\end{equation}
where the redefined couplings are
\begin{eqnarray}
\tilde{g}^2&=&g^2(1+\mu \eta+\bar \mu \bar{\eta}+2\mu\bar \mu \eta
\bar{\eta}),\ \ \  \eta=\theta^2, \ \ \ \bar{\eta}=\bar{\theta}^2,
\label{g}\\
\tilde{\lambda}^{ijk}&=&\lambda^{ijk}-A^{ijk}\eta +\frac 12
(\lambda^{njk}(m^2)^i_n +\lambda^{ink}(m^2)^j_n+\lambda^{ijn}(m^2)^k_n)\eta
\bar \eta,  \label{y1}\\
\tilde{\bar \lambda}_{ijk}&=&\bar \lambda_{ijk} -
\bar A_{ijk} \bar{\eta}+ \frac 12 (\bar \lambda_{njk}(m^2)_i^n
+\bar \lambda_{ink}(m^2)_j^n+\bar \lambda_{ijn}(m^2)_k^n)\eta \bar \eta
 .  \label{y2}
\end{eqnarray} }
\end{statement}

From eqs.(\ref{Z}) and (\ref{g}-\ref{y2}) it is possible to write down
explicit differentials operators which have to be applied to the
$\beta $ functions of a rigid theory in order to get those for the soft
terms.

Relations between the rigid and soft terms renormalizations are summarized
in the Table.
\begin{center}
\begin{tabular}{|l|l|}
\hline \hline  & \\[-0.2cm]
\hspace*{1.5cm}  The Rigid Terms & \hspace*{1.5cm}  The Soft Terms \\[0.2cm]
\hline  & \\
$\beta_{\alpha_i} =  \alpha_i\gamma_{\alpha_i}$ &
$\beta_{m_{A i}}=D_1\gamma_{\alpha i}$ \\[0.3cm]
$\beta_{M}^{ij} =\frac{1}{2}(M^{il}\gamma^j_l+M^{lj}\gamma^i_l) $&
$\beta_{B}^{ij} = \frac{1}{2}(B^{il}\gamma^j_l+B^{lj}\gamma^i_l)-
(M^{il}D_1\gamma^j_l+M^{lj}D_1\gamma^i_l) $\\[0.3cm]
$\beta_{y}^{ijk} =\frac{1}{2}(y^{ijl}\gamma^k_l+y^{ilk}\gamma^j_l+
y^{ljk}\gamma^i_l)$ &
$\beta_{A}^{ijk} = \frac{1}{2}(A^{ijl}\gamma^k_l+A^{ilk}\gamma^j_l+
A^{ljk}\gamma^i_l)$ \\[0.2cm] &
\hspace*{1cm}
$-(y^{ijl}D_1\gamma^k_l+y^{ilk}D_1\gamma^j_l+y^{ljk}D_1\gamma^i_l)$\\[0.2cm]
&  $(\beta_{m^2})^i_j=D_2\gamma^i_j $ \\[0.3cm]
\hline
\multicolumn{2}{|l|}  {} \\[-0.2cm]
\multicolumn{2}{|l|}
{$D_1= m_{A_i}\alpha_i\frac{\displaystyle \partial}{\displaystyle \partial
\alpha_i} -A^{ijk}\frac{\displaystyle \partial}{\displaystyle \partial
y^{ijk}}\ \ , \hspace{1.7cm}
\bar{D}_1=m_{A_i}\alpha_i\frac{\displaystyle  \partial}{\displaystyle
\partial \alpha_i} -A_{ijk}\frac{\displaystyle \partial}{\displaystyle
\partial y_{ijk}} $ }\\[0.3cm]
\multicolumn{2}{|l|} {$D_2= \bar{D}_1 D_1 +
m_{A_i}^2\alpha_i\frac{\displaystyle \partial }{\displaystyle \partial
\alpha_i} $} \\[0.3cm] \multicolumn{2}{|l|} {\hspace*{0.1cm}
$+\frac{1}{2}(m^2)^a_n\left(y^{nbc}\frac{\displaystyle \partial
}{\displaystyle \partial y^{abc}} +y^{bnc}\frac{\displaystyle \partial
 }{\displaystyle \partial y^{bac}}+ y^{bcn}\frac{\displaystyle \partial
}{\displaystyle \partial y^{bca}}+ y_{abc}\frac{\displaystyle \partial
}{\displaystyle \partial y_{nbc}}+ y_{bac}\frac{\displaystyle \partial
}{\displaystyle \partial y_{bnc}}+ y_{bca}\frac{\displaystyle \partial
}{\displaystyle \partial y_{bcn}}\right)$ }\\[0.3cm] \hline \hline
\end{tabular}
\end{center}

\noindent where to simplify the formulae, we use the following
notation:
$$\alpha_i = g^2_i/16\pi^2, \ \ y^{ijk}=\lambda^{ijk}/4\pi,
\ \ y_{ijk}=\bar \lambda_{ijk}/4\pi, \ \ A^{ijk}=A^{ijk}/4\pi, \ \
 A_{ijk}=\bar A_{ijk}/4\pi. $$

\section{Renormalization of the Soft Terms in SUSY GUTs}

The general rules described in the previous section can be applied to any
model, in particular to a SUSY GUT. In the case when the field
content and the Yukawa interactions are fixed, it is more useful to deal with
numerical rather than with tensor couplings. Rewriting the
superpotential (\ref{rigid}) and the soft terms (\ref{sofl}) in terms
of group invariants, one has
\begin{equation}
{\cal W}_{SUSY}=\frac{1}{6}\sum_a y_a\lambda^{ijk}_a
\Phi_i\Phi_j\Phi_k+\frac{1}{2}\sum_b M_bh^{ij}_b\Phi_i\Phi_j, \label{pot}
\end{equation}
and
\begin{equation} -{\cal L}_{soft}=\left[\frac 16 \sum_a {\cal A}_a
\lambda^{ijk}_a \phi_i\phi_j\phi_k+ \frac 12 \sum_b {\cal
B}_bh^{ij}_b\phi_i\phi_j + \frac 12 m_{A_j}\lambda_j\lambda_j+h.c.\right]
+(m^2)^j_i\phi^{*i}\phi_j,\label{sof}
\end{equation}
where we have introduced numerical couplings $y_a,M_b,{\cal A}_a$ and
${\cal B}_b$.

Usually, it is assumed that the soft terms obey the universality
hypothesis, i.e. they repeat the structure of a superpotential, namely
\begin{equation}
{\cal A}_a=y_aA_a,\ \ {\cal B}_b=M_bB_b, \ \ (m^2)^i_j=m^2_i\delta^i_j.
\end{equation}
Thus, we have the following set of couplings and soft parameters:
$$
g_j,\ y_a,\ M_b,\ A_a,\ B_b, \ m^2_i,\ m_{A_j}.
$$
Then, the renormalization group $\beta$ functions of a rigid theory
(\ref{rigidlag},\ref{rigid}) look like (we assume the
diagonal renormalization of matter superfields)
\begin{eqnarray}
\beta_{\alpha_j} &=& \beta_j \equiv \alpha_j\gamma_{\alpha_j}, \\
\beta_{y_a} &=& \frac{1}{2}y_a \sum_iK_{ai}\gamma_i, \ \ \
\sum_iK_{ai}=3, \\
\beta_{M_b} &=& \frac{1}{2}M_b\sum_iT_{bi}\gamma_i,\ \ \
\sum_iT_{bi}=2,
\end{eqnarray}
where $\gamma_i$ is the  anomalous dimension of the superfield
$\Phi_i$, $\gamma_{\alpha_j}$ is the anomalous dimension of the gauge
superfield (in some gauges) and numerical matrices $K$ and $T$ specify
which particular fields contribute to a given term in eq.(\ref{pot}).

To get the renormalization of the soft terms, one has to apply the
formulae of the previous section. In terms of numerical couplings they are
simplified.

The renormalizations of the soft terms are expressed through those
of a rigid theory in the following way:
\begin{eqnarray}
\beta_{m_{A_j}}&=&D_1\gamma_{\alpha_j}, \label{ma}\\
\beta_{A_a}&=&-D_1\sum_i K_{ai}\gamma_i, \label{A}\\
\beta_{B_b}&=&-D_1\sum_i T_{bi}\gamma_i, \label{B}\\
\beta_{m^2_i}&=&D_2\gamma_i,  \label{m2}
\end{eqnarray}
and the operators $D_1$ and $D_2$ now take the form
\begin{eqnarray}
D_1&=&m_{A_i}\alpha_i \frac{\partial }{\partial \alpha_i}
 -A_aY_a\frac{\partial }{\partial Y_a},  \label{d1}\\
D_2&=&(m_{A_i}\alpha_i \frac{\partial }{\partial \alpha_i}
 -A_aY_a\frac{\partial }{\partial Y_a})^2
+m_{A_i}^2\alpha_i\frac{\partial }{\partial \alpha_i}+
m^2_iK_{ai}Y_a\frac{\partial }{\partial Y_a}. \label{d2}
\end{eqnarray}
where $Y_a=y_a^2.$

\section{Finiteness of Soft Parameters in a Finite SUSY GUT}

Consider now the application of the proposed formulae to construct
totally finite softly broken theories.

For rigid N=1 SUSY theories there exists a general method of
constructing totally all loop finite gauge theories  proposed in
refs.~\cite{EKT, Jones1, Sibold}.  The key issue of the method is the
one-loop finiteness. If the theory is one-loop finite and satisfies
some criterion verified in one loop~\cite{Kazakov}, it can be
made finite in any loop order by fine-tuning of the Yukawa couplings
order by order in PT. In case of a simple gauge group the Yukawa
couplings have to be chosen in the form
\begin{equation}
Y_a(\alpha)=c^a_1\alpha +c^a_2\alpha^2+..., \label{yuk}
\end{equation}
where the finite coefficients  $c^a_n$ are calculated algebraically in
the n-th order of perturbation theory. Since the one-loop finite theory
is automatically two-loop finite~\cite{Finite}, the coefficients
$c^a_2=0$.

Suppose now that a rigid theory is made finite to all orders by the
choice of the Yukawa couplings as in eq.(\ref{yuk}).
This means that all the anomalous dimensions and the $\beta $ functions
on the curve $Y_a=Y_a(\alpha)$ are identically equal to zero\footnote{
Sometimes by a finite theory it is understood that all $\beta $
functions vanish but not necessarily the anomalous dimensions
$\gamma_i$. However, there is no big difference and one can reach
vanishing of all $\gamma_i$ as well~\cite{Kazakov}.}.

Consider the renormalization of the soft terms.  According to
eqs.(\ref{ma}-\ref{m2}) and (\ref{d1},\ref{d2}) the renormalizations of
the soft terms are not independent but are given by the differential
operators acting on the same anomalous dimensions.  One has either
\begin{equation}
\beta_{soft} \sim D_1\gamma (Y,\alpha )=( m_A\alpha
\frac{\partial }{\partial \alpha }- A_aY_a\frac{\partial }{\partial
Y_a} )\gamma (Y_a,\alpha ), \label{b1}
\end{equation}
or
\begin{equation}
\beta_{soft} \sim D_2\gamma (Y,\alpha )=[( m_A\alpha
\frac{\partial }{\partial \alpha }- A_aY_a\frac{\partial }{\partial
Y_a})^2+ m_A^2\alpha \frac{\partial }{\partial \alpha }+
m_i^2K_{ai}Y_a\frac{\partial }{\partial Y_a}]\gamma (Y_a,\alpha ),
\label{b2}
\end{equation}
where  $ \gamma (Y_a,\alpha ) $ is some anomalous dimension.

From the requirement of finiteness
\begin{equation}
\gamma_i(Y_a(\alpha),\alpha )=0,
\end{equation}
the Yukawa couplings $Y_a(\alpha )$ are found in the form (\ref{yuk}).

To reach the finiteness of all the soft terms in all
loop orders, one has to choose the soft parameters
$A_a$ and $m^2_i$ in a proper way. The following statement is valid:

\begin{statement}
{\it The soft term $\beta $ functions become equal to zero
if the parameters $A_a$ and $m^2_i$ are chosen in the following form:
\begin{eqnarray} A_a(\alpha )&=&-m_A\alpha
\frac{\partial }{\partial \alpha }\ln Y_a(\alpha ), \label{fina}\\
m^2_i&=&-m_AK^{-1}_{ia}\frac{\partial }{\partial \alpha }\alpha A_a(\alpha )
\label{finm}          \\
&=&m_A^2K^{-1}_{ia}\frac{\partial }{\partial \alpha }\alpha^2
\frac{\partial }{\partial \alpha }\ln Y_a(\alpha ), \nonumber
\end{eqnarray}
where the matrix $K^{-1}_{ia}$ is the inverse of the matrix $K_{ai}$.}
\end{statement}
This statement follows from the form of the operators $D_1$ and $D_2$. After
substitution of solutions (\ref{fina}) and (\ref{finm}) into $D_1$
and $D_2$, $D_1$ becomes a total derivative over $\alpha $ and $D_2$
becomes a second total derivative.

Indeed, consider eq.(\ref{b1}). For $A_a$ chosen as in eq.(\ref{fina}) the
differential operator $D_1$ takes the form
$$D_1=m_A\alpha \frac{\partial }{\partial \alpha }-
A_aY_a\frac{\partial }{\partial Y_a}
=m_A(\frac{\partial \ln  Y_a}{\partial \ln \alpha }
\frac{\partial }{\partial \ln Y_a} +
\frac{\partial }{\partial \ln \alpha })=m_A\frac{d}{d\ln \alpha }.$$
Hence, since on the curve $Y_a=Y_a(\alpha )$ the anomalous dimension
$\gamma (Y_a,\alpha )$ identically vanishes, so does its derivative
$$\frac{d}{d\ln \alpha }\gamma (Y_a(\alpha ),\alpha ) =0.$$
The  situation  with the operator $D_2$ is a bit more complicated. In this
case, one has the second derivative.  However, using eq.(\ref{finm})
one has
\begin{eqnarray*}
D_2&=&(m_A\alpha \frac{\partial }{\partial \alpha }-
A_aY_a\frac{\partial }{\partial Y_a})^2+ m_A^2\alpha \frac{\partial
}{\partial \alpha }+ m_i^2K_{ai}Y_a\frac{\partial }{\partial Y_a}\\
&=& (m_A\alpha \frac{\partial }{\partial \alpha }- A_aY_a\frac{\partial
}{\partial Y_a})^2 - m_A\frac{\partial A_a}{\partial \ln \alpha }
Y_a\frac{\partial }{\partial Y_a} +m_A( m_A\alpha \frac{\partial
}{\partial \alpha }- A_aY_a\frac{\partial }{\partial Y_a})\\
& =& m_A\frac{d}{d \ln \alpha }+m_A^2\frac{d^2}{d\ln^2 \alpha}.
\end{eqnarray*}
The term with the derivative of $A_a$ is essential to
get the total second derivative over $\alpha $, since in the bracket
the derivative $\alpha  \partial/\partial \alpha $  does not act on $A_a$
by construction.

Like in the previous case the total derivatives identically
vanish on the curve $Y_a=Y_a(\alpha )$
$$(m_A\frac{d}{d\ln \alpha }+m_A^2\frac{d^2}{d\ln^2 \alpha })\gamma
 (Y_a(\alpha ),\alpha ) =0.$$

The solutions (\ref{fina},\ref{finm}) can be checked perturbatively
order by order. It is not difficult to see the general combinatoric law
and to check their validity. We have done this in two
particular cases: a general finite SUSY theory with one Yukawa coupling and
the finite SUSY SU(5) GUT with 5 Yukawa couplings considered in
ref.~\cite{EKT}. In both ther cases the calculation has been performed
up to three loops.

In the leading order eqs.(\ref{fina}, \ref{finm}) give
\begin{equation}
A_a=-m_A,  \ \ \ m^2_i=\frac{1}{3}m_A^2,   \label{oneloop}
\end{equation}
since $\sum_aK^{-1}_{ia}=1/3$. These relations coincide with the already
known ones~\cite{Mezinchescu} and with those coming from
supergravity~\cite{Nilles} and supersting-inspired
models~\cite{Ibanez}. There  they  usually
follow from the requirement of finiteness of the cosmological
constant and probably have the same origin. Note that since the
one-loop finiteness of a rigid theory automatically leads to the
two-loop one and hence the coefficients $c^a_2=0$, the same statement
is valid due to eqs.(\ref{fina},\ref{finm}) for a softly broken theory.
Namely, relations (\ref{oneloop}) are valid up to two-loop order
in accordance with ~\cite{Jones}.
In higher orders, however, they have to be modified.

The parameters $M_b$ and $B_b$ are not specified and are finite
provided eqs.(\ref{fina}),(\ref{finm}) are satisfied. In particular GUT
models, some of parameters $B_b$ may be equal to $A_b$ for the reason of
fine-tuning during spontaneous breaking of a GUT symmetry group.

\section{Conclusion}

Thus, we have demonstrated how one can construct all loop finite
N=1 SUSY GUT including the soft SUSY breaking terms. Remarkably
that in the leading order the relations between the soft
terms  coincide with those of supergravity and superstring theory. In
higher orders, however, one needs some fine-tuning which is given by
exactly the same functions as in a rigid theory.

\vspace{1cm}

{\large \bf Acknowledgements}

\vspace{0.3cm}

The author is grateful to I.Kondrashuk and A.Gladyshev for
valuable discussions.  Financial support from RFBR
grants \# 96-02-17379a and \# 96-15-96030 is kindly acknowledged.

\end{document}